\documentclass[aps,prb,twocolumn,groupedaddress,showpacs]{revtex4}

\usepackage{graphicx}
\usepackage{dcolumn}
\usepackage{bm}
\begin{document}

\title{Evolution of the resistivity anisotropy in 
Bi$_{2}$Sr$_{2-x}$La$_x$CuO$_{6+\delta}$ single crystals 
for a wide range of hole doping}

\author{S. Ono}
\author{Yoichi Ando}
\email[]{ando@criepi.denken.or.jp}
\affiliation{
Central Research Institute of Electric Power Industry, Komae, Tokyo
201-8511, Japan}

\date{\today}

\begin{abstract}

To elucidate how the temperature dependence of the resistivity
anisotropy of the cuprate superconductors changes with hole doping, both
the in-plane and the out-of-plane resistivities ($\rho_{ab}$ and
$\rho_c$) are measured in a series of high-quality
Bi$_{2}$Sr$_{2-x}$La$_x$CuO$_{6+\delta}$ (BSLCO) single crystals for a
wide range of $x$ ($0.23 \le x \le 1.02$), which corresponds to the hole
doping per Cu, $p$, of 0.03 -- 0.18. The anisotropy ratio,
$\rho_{c}/\rho_{ab}$, shows a systematic increase with decreasing $p$ at
moderate temperatures, except for the most underdoped composition where
the localization effect enhances $\rho_{ab}$ and thus lowers
$\rho_{c}/\rho_{ab}$. The exact $p$ dependence of
$\rho_{c}/\rho_{ab}$ at a fixed temperature is found to be quite
peculiar, which is discussed to be due to the effect of the pseudogap
that causes $\rho_{c}/\rho_{ab}$ to be increasingly more enhanced as
$p$ is reduced. The pseudogap also causes a rapid
growth of $\rho_{c}/\rho_{ab}$ with decreasing temperature, and, as a
result, the $\rho_{c}/\rho_{ab}$ value almost reaches 10$^{6}$ in
underdoped samples just above $T_c$. Furthermore, it is found that the
temperature dependence of $\rho_c$ of underdoped samples show two
distinct temperature regions in the pseudogap phase, which suggests
that the divergence of $\rho_c$ below the pseudogap temperature is 
governed by two different mechanisms.

\end{abstract}


\maketitle

\section{INTRODUCTION}
It has long been recognized that the peculiar $c$-axis transport 
properties \cite{Gray} 
of the high-$T_c$ cuprates are strong manifestations of 
the unusual electronic state in these materials. \cite{Anderson} 
In particular, the following two features have been considered to be 
most unusual:
(i) The magnitude of the $c$-axis resistivity $\rho_c$ is orders of 
magnitude larger than that expected from band calculations, \cite{Pickett} 
leading to a huge resistivity anisotropy where $\rho_c$ is up to 10$^5$ 
times larger \cite{Martin} than the in-plane resistivity $\rho_{ab}$. 
(ii) The temperature dependence of $\rho_c$ is in most cases 
semiconducting or insulating (i.e., $d\rho_c/dT < 0$), while that of 
$\rho_{ab}$ is metallic ($d\rho_{ab}/dT > 0$); 
such a contrasting behavior \cite{Nakamura,Takenaka,Ando} 
is not expected in ordinary anisotropic metal. \cite{Gray,Anderson}

The feature (i) indicates that the strong correlations in the 
cuprate materials give rise to some unconventional mechanism 
which renormalizes the $c$-axis transfer-matrix element to a value 
much smaller than what is expected for an uncorrelated system.
This renormalization of the $c$-axis transfer-matrix element is 
usually called ``charge confinement".  
There have been many theoretical proposals to explain the 
charge confinement in the cuprates, but it is fair to say that the 
confinement mechanism is not yet understood on the fundamental level.  
Since many of the theoretical models for the charge confinement, 
such as the resonating-valence-bond theories \cite{Zou,Nagaosa} or 
self-organized stripe theories, \cite{Carlson,Zaanen} are closely 
tied to the mechanism of the high-$T_c$ 
superconductivity, it remains to be important to study the $c$-axis 
transport in the cuprates. 

The feature (ii) is now better understood than the feature (i), 
both experimentally and theoretically, though the understanding 
is still far from complete. 
To appreciate the current understanding of the feature (ii), 
it should first be recognized \cite{Ioffe} that the $c$-axis transport 
in the normal-state of the cuprates is essentially an incoherent tunneling 
process; this fact gives rise to the intrinsic Josephson effect 
\cite{Muller} as well 
as the Josephson-plasma resonance \cite{JPR} in the superconducting state. 
It should at the same time be noted that the tunneling conductance 
for ordinary metal-insulator-metal junctions should be 
\textit{independent} of temperature, according to the Fermi golden rule; 
this means that the ``semiconducting" behavior of $\rho_c(T)$ 
does \textit{not} 
automatically arise from the tunneling nature of the $c$-axis transport. 
(In this sense, the expression ``semiconducting behavior", which implies 
that the transport is essentially an activation process over a gap, 
is inappropriate, though this expression is often naively 
used in this context.) 
Thus, there must be some unconventional mechanisms which cause the peculiar 
``insulating" temperature dependence of $\rho_c$ in the cuprates.

Experimentally, there seem to be two additive mechanisms that 
are both responsible for the ``insulating" $\rho_c(T)$. 
The first is the charge-confinement mechanism which appears to 
become increasingly more effective with lowering temperature; 
this causes the tunneling matrix element to be reduced and keeps $\rho_c$ 
increasing with decreasing $T$.  This point was most clearly 
demonstrated \cite{Ando} by the low-temperature normal-state resistivity 
measurement 
of Bi$_{2}$Sr$_{2-x}$La$_x$CuO$_{6+\delta}$ (BSLCO) 
under 60 T, which showed that $\rho_c$ of Bi-2201 keeps increasing 
even down to 0.6 K. 
The more recently recognized mechanism is the effect of the pseudogap 
\cite{Timusk} which causes destruction of the Fermi surface starting 
from the ($\pm \pi$, 0), (0, $\pm \pi$) points, as has been demonstrated 
by angle-resolved photoemission spectroscopy (ARPES) 
measurements \cite{Norman}; 
this causes the available density of states (DOS) for tunneling to be reduced, 
and $\rho_c$ gets steeply increased as the pseudogap develops.
Note that the $c$-axis matrix element of the cuprates has a particular 
$\mathbf{k}$-dependence \cite{Xiang} 
($\mathbf{k}$ is the in-plane wave vector of the conduction electrons) 
that tends to amplify the contribution 
of the electrons on these gapped portion of the Fermi surface, 
and this $\mathbf{k}$-dependence of the matrix element causes $\rho_c$ 
to be very sensitive to the opening of the pseudogap. \cite{Ioffe}
Recent tunneling-spectroscopic studies \cite{Suzuki,Krasnov} 
of the intrinsic junctions in 
Bi$_{2}$Sr$_{2}$CaCu$_{2}$O$_{8+\delta}$ (Bi-2212) have made strong 
cases for the interpretation that the steep upturn in $\rho_c(T)$ below 
the pseudogap temperature $T^*$ is largely due to this second mechanism. 
It is expected that these two independent mechanisms, confinement and 
pseudogap, together cause the rather complicated, ``insulating" 
$\rho_c(T)$ behavior in the cuprates, though their relative roles in 
determining the actual temperature dependence is not clarified yet.

Given the above-mentioned recent understanding of the $c$-axis transport, 
it would be instructive to see how 
the two mechanisms (confinement and pseudogap) are actually working in a 
single system. 
For such a purpose, it is useful to elucidate the evolution of the behavior 
of $\rho_c(T)$, as well as the anisotropy ratio $\rho_c/\rho_{ab}$, over 
a wide range of hole doping, because $T^*$ is known to decreases with 
increasing doping.  
In the past, there were many studies of the anisotropic resistivities 
in La$_{2-x}$Sr$_x$CuO$_4$ (LSCO), \cite{Nakamura,Kimura,logT,Boebinger} 
YBa$_{2}$Cu$_3$O$_{7-\delta}$ (YBCO), \cite{Takenaka,Terasaki} 
Bi-2212, \cite{Watanabe97} and BSLCO, \cite{Martin,Wang} 
but none of them cover a wide enough doping range from 
underdoped insulators to overdoped superconductors to yield 
a coherent picture. 

Very recently, Komiya \textit{et al.} measured \cite{Komiya} the anisotropic 
resistivity of high-quality LSCO single crystals for $x$ = 0.01 -- 0.10 
and found that in this system the effect of pseudogap is {\it not} apparent 
(i.e., the additional mechanism to cause the insulating $\rho_c(T)$ is 
absent), probably because of the unique Fermi surface topology 
\cite{Zhou,InoFS} of LSCO. 
Therefore, it is useful to study another cuprate system in which one 
can change the hole doping over a really wide range.  Obviously, BSLCO 
is an ideal candidate, because we have demonstrated 
\cite{Murayama,Ono,Hanaki} that the hole 
doping can be varied (by changing the La concentration) from 
heavily-underdoped insulator to overdoped superconductor; moreover, 
ARPES measurements have found \cite{Harris,Chuang,Sato} that 
the Fermi surface topology and the 
pseudogap structure of BSLCO are very similar to those of Bi-2212. 
We can thus expect that it may be possible to substantiate the  
effects of charge confinement and the pseudogap on the $c$-axis transport 
by tracing its evolution with the hole doping in BSLCO.

The purpose of this paper is to provide high-quality data, 
obtained from well-controlled experiments, on the 
evolution of the temperature dependences of $\rho_c$ and 
$\rho_c/\rho_{ab}$ of BSLCO for a wide range of doping, which showcases 
the complicated origin of the unusual $c$-axis transport in the cuprates. 
Our data exemplifies that there are three different mechanisms that 
govern the behavior of the anisotropy ratio $\rho_c/\rho_{ab}$: 
charge confinement, pseudogap, and localization. 
In particular, the temperature dependence of $\rho_c$ of our underdoped 
samples makes a good case for the two different origins of the 
insulating behavior below $T^*$.

This paper is organized as follows: 
Details of our crystal-growth technique to produce the high-quality 
BSLCO single crystals, as well as the details of our measurement technique, 
are presented in Section II.  
Section III presents our data for $\rho_{ab}(T)$, $\rho_c(T)$, and 
$\rho_{c}/\rho_{ab}$, putting emphasis on how they evolve with 
changing hole doping.  The implication of the systematics of our observation 
is discussed in Section IV, and Section V summarizes our findings.

\section{EXPERIMENTS}

\subsection{Bi$_{2}$Sr$_{2-x}$La$_x$CuO$_{6+\delta}$ Crystals}

The Bi$_{2}$Sr$_{2-x}$La$_x$CuO$_{6+\delta}$ system is a Bi-based 
single-layer cuprate where the hole 
doping is controlled by replacing Sr with La.  Until a few years ago, 
this material had been considered to be a particularly ``dirty" system 
among the cuprates, 
because the in-plane resistivity of even the best crystal at those 
times showed the residual resistivities of around 70 $\mu\Omega$cm 
(Refs. \onlinecite{Martin,Ando}) and 
the Hall coefficient $R_H$ displayed only a weak temperature dependence, 
\cite{Mackenzie}
both of which are characteristic of dirty cuprate samples. 

In 1999, Ando and Murayama demonstrated \cite{Murayama} 
that it is possible to grow 
high-quality single crystals of BSLCO that show much smaller residual 
resistivity and a strongly temperature-dependent $R_H$, which conform to 
the ``standard" behavior of cuprates.  Also, though the BSLCO system was 
believed to possess the lowest optimum $T_c$ among the well-studied 
cuprate systems, we have demonstrated \cite{Ono} 
that the optimum $T_c$ of this system can be raised to 38 K, 
which is almost equal to the optimum $T_c$ of LSCO.  

In BSLCO, larger La concentration $x$ corresponds to smaller 
hole doping per Cu, $p$, and the relation between $x$ and $p$, as well as 
the phase diagram of $T_c$ vs $p$, has been sorted out \cite{Hanaki}; 
it was found 
that optimum doping occurs at $p \simeq$ 0.16 (which is achieved with 
$x \simeq$ 0.4), but the superconductivity disappears at $p \simeq$ 0.10 
($x \simeq$ 0.8) upon underdoping.  Low-temperature normal-state 
resistivity measurement of this system using 60-T pulsed magnetic 
field found \cite{Ono} that the metal-to-insulator crossover occurs at 
$p \simeq$ 1/8, below which $\rho_{ab}(T)$ shows a peculiar $\log (1/T)$ 
divergence.

Single crystals of Bi$_{2}$Sr$_{2-x}$La$_x$CuO$_{6+\delta}$ are  
grown by the floating-zone (FZ) technique for a wide range of La 
concentration (0.23 $\le x \le$ 1.02). 
Before the FZ operation, we first prepare polycrystalline rods of BSLCO. 
Raw powders of Bi$_2$O$_3$, SrCO$_3$, La$_2$O$_3$, and CuO, with purities 
of 99.9\% or higher, are dried, weighed, mixed into the nominal molar 
ratio of the target composition, and well ground in an agate mortar; 
they are then calcined at 750 -- 850$^\circ$C for 20 hours to form 
the BSLCO phase. 
The resulting powders are reground and calcined again, and this process 
is repeated twice. 
The x-ray diffraction analysis reveals that the powders become 
100\%-pure BSLCO phase after the third calcination. 
The resulting BSLCO powders are isostatically pressed into a rod shape 
($\approx$ 5 mm$\phi$ $\times$ 100 mm) and finally sintered at 
850$^\circ$C for 20 hours, to form a rigid polycrystalline feed rod to 
be used for the FZ operation. 

Single-crystal growth is carried out using an infrared image furnace 
(NEC Machinery SC K-15HD) with two halogen lamps and double ellipsoidal 
mirrors.  We grow BSLCO crystals with a ``simple" FZ technique, 
which means that we do not use any solvent to reduce the temperature of 
the molten zone. 
One important technique we use is what we call ``fast scan", with which 
we quickly melt and quench the whole feed rod using the FZ furnace 
with the feed speed of 30 mm/h.  This process is necessary for making 
the feed rod dense, which is important for avoiding the 
emergence of bubbles during the crystal growth; actually, the rod 
becomes appreciably thinner than the as-sintered rod 
after the fast scan (it becomes roughly 3 mm$\phi$).
The ``scanned" rod is then remounted to the FZ furnace and is used 
for the actual crystal growth.  The growth rate is kept constant at 
0.5 mm/h, which is two times slower than the growth rate 
for LSCO. \cite{Komiya}
Another important technique we use during the growth is what we call 
``necking"; since many independent domains start to crystallize at the 
initial stage of the growth, after roughly 1 cm of the rod is grown, 
the diameter of the molten zone is gradually reduced to 
approximately 2 mm$\phi$, which causes fewer domains to remain 
in the crystal.  The molten zone is then gradually fattened, which 
leads to the growth of the fewer domains in the resulting rod. 
After a few times of the necking, when the growth is successful, 
there remains only one domain and the whole chunk becomes a large 
single-domain crystal.
The space around the molten zone is filled with 1 atm of flowing oxygen 
during the growth of underdoped crystals, while for the growth of 
overdoped crystals the space is filled with 1 atm of flowing air.
The size of the single-domain crystals that can be obtained from 
the grown rod is typically 20 $\times$ 3 mm$^2$. 

\begin{table}
\caption{Actual composition of the obtained crystals for each 
nominal (starting) La concentration.  
The molar ratios of the cations are determined 
by ICP-AES and are shown with the Cu value fixed to 1.}
\begin{ruledtabular}
\begin{tabular}{ccccc}
nominal La composition& Bi& Sr& La& Cu\\
\hline
0.20& 2.02& 1.70& 0.23& 1\\
0.40& 1.98& 1.59& 0.39& 1\\
0.50& 1.99& 1.52& 0.49& 1\\
0.60& 1.93& 1.31& 0.66& 1\\
0.70& 2.00& 1.30& 0.73& 1\\
0.80& 2.00& 1.17& 0.84& 1\\
0.90& 1.92& 1.06& 0.92& 1\\
1.00& 1.90& 0.98& 1.02& 1\\
\end{tabular}
\end{ruledtabular}
\end{table}

The actual La concentrations in the 
crystals are determined by the inductively-coupled plasma 
atomic-emission spectroscopy (ICP-AES).  
The crystals tend to contain more La than the nominal composition, 
but the difference is small; here we report crystals with 
actual $x$ values of 0.23, 0.39, 0.49, 0.66, 0.73, 0.84, 0.92, and 1.02, 
and these crystals were obtained from the nominal $x$ values of 
0.20, 0.40, 0.50, 0.60, 0.70, 0.80, 0.90, and 1.00, respectively. 
In all these crystals, the ICP-AES analysis confirms that there is 
essentially no excess Bi in the crystals; rather, Bi tends to be 
deficient in highly La-doped crystals, as shown in Table I. 
On the other hand, the ratio of (Sr+La):Cu is always near 2:1.

Uniformity of the distribution of the La dopants in the 
crystals is confirmed by the electron-probe microanalysis (EPMA), 
with which the variation of the $x$ value is estimated to be within 5\%. 
The superconducting transition temperature $T_{c}$ is determined both 
by the zero resistivity in the transport measurement and by the onset 
of the Meissner signal in the SQUID magnetization measurement; for all 
the crystals reported here, these two definitions give $T_c$ values 
that match within $\pm$0.5 K, and the variation among different pieces 
of crystals with the same composition is also within $\pm$0.5 K. 
(Thus we refer to the $T_c$ values with two significant digits in this 
paper.) 
The $T_c$ values for each of the compositions are listed in Table II.

\begin{table}
\caption{Actual hole concentrations per Cu, $p$, 
and $T_c$ for each La concentration $x$.  The $p$ values are determined 
from the empirical relation between $x$ and $p$ obtained in Ref. 
\onlinecite{Hanaki}.}
\begin{ruledtabular}
\begin{tabular}{ccccccccc}
 $x$& 0.23& 0.39& 0.49& 0.66& 0.73& 0.84& 0.92& 1.02\\
\hline
$p$& 0.18& 0.16& 0.14& 0.12& 0.11& 0.10& 0.07& 0.03\\
$T_c$& 29& 38& 31& 23& 14& 1.4& 0& 0\\
\end{tabular}
\end{ruledtabular}
\end{table}

\subsection{Measurements}

The crystals are cut into dimensions of 
typically 2 $\times$ 1 $\times$ 0.05 mm$^3$ for the $\rho_{ab}$ 
measurements and 1 $\times$ 1 $\times$ 0.05 mm$^3$ for the $\rho_{c}$ 
measurements. 
Accurate determination of the crystal thickness is done by 
measuring the weight with 0.1 $\mu$g resolution and converting it 
into volume using the nominal density for each composition. 
We employ the straightforward four-terminal method to measure 
$\rho_{ab}$ and $\rho_c$:  
For the $\rho_{ab}$ measurements, the current contacts are painted to 
cover two opposing side faces of the platelet-shaped crystals to 
ensure a uniform current flow and the voltage contacts are painted on 
the two remaining side faces of the crystals.  For the $\rho_{c}$ 
measurements, two current contacts are painted to almost completely cover 
the opposing $ab$ faces of the crystal and two voltage contacts are placed 
in the small window reserved in the center of the current contacts. 
In both cases, the contact pads are tactfully hand-drawn with gold paint, 
followed by a heat treatment at 400$^{\circ}$C for 30 minutes in the air, 
which makes the gold particles to well adhere to the sample surface.
After this heat-treatment to cure the gold contact pads, the samples are 
annealed at higher temperatures for longer time to control the oxygen 
concentration; samples with $x$ larger than 0.3 are annealed in flowing 
air at 650$^\circ$C for 48 hours, while those with $x \le 0.3$ 
are annealed in flowing oxygen at 400$^\circ$C for 60 hours. 
In both cases, the crystals are annealed together with a sufficient amount 
of BSLCO powders of the same composition, and they are quenched to 
room temperature at the end of the annealing.  
Finally, gold lead wires are attached to the contact pads using silver epoxy, 
which is cured at relatively low temperature, 130$^{\circ}$C. 
The contact resistance we achieve with this technique is less than 1 $\Omega$.
We note that the above annealing conditions are chosen to minimize the 
superconducting transition width $\Delta T_c$ of our crystals; 
in all the superconducting samples, $\Delta T_c$ in $\rho_c(T)$ data is 
around 2 K after the above annealing. 

\begin{figure}
\includegraphics[width=8cm]{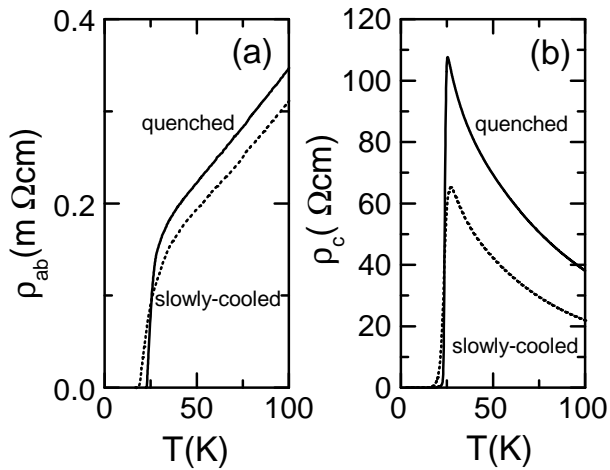}
\caption{Temperature dependences of (a) $\rho_{ab}$ and (b) $\rho_c$ of 
the BSLCO crystals with $x$ = 0.66 ($p$ = 0.12); 
quenched samples (solid lines) and 
slowly-cooled samples (dotted lines) show non-negligible difference, 
even though the annealing conditions are the same.}
\end{figure}

Before finishing the description of our sample preparation technique, 
we would like to emphasize the importance of the quenching at the end 
of the high-temperature annealing. 
Figure 1 shows the data of $\rho_{ab}(T)$ and $\rho_c(T)$ 
of our BSLCO crystals with $x$ = 0.66, for both quenched 
and slowly-cooled samples.  Although both sets of the samples are 
annealed at 650$^\circ$C for 48 hours, the data show non-negligible 
differences; the difference in the $\rho_{ab}(T)$ are comparatively 
small (the magnitude of $\rho_{ab}$ of quenched sample is 10\% larger 
than that of slowly-cooled sample), but the difference in $\rho_{c}(T)$ 
is significant, differing by about 50\%.  
Also, $\Delta T_{c}$ is about 2 K for the quenched samples, 
while it is about 6 K for the slowly-cooled sample. 
Those differences are consistent with the interpretation 
that the slow cooling causes the oxygen concentration in the samples 
to be inhomogeneous and there are more oxygen near the surface of 
the slowly-cooled samples; remember that the equilibrium oxygen 
concentration of Bi-2201 during the annealing is dependent on both the 
temperature and the oxygen partial pressure, and the equilibrium 
concentration becomes higher as the temperature is lowered. 
Thus, to obtain uniformly oxygenated samples and reproducible data, 
quenching is indispensable. 

In this work, we measure $\rho_{ab}$ and $\rho_{c}$ on more than 20 
samples for each La concentration to check for the reproducibility; 
the variation in the absolute magnitude of the resistivity of the 
quenched sample is less than $\pm$10\%, which is consistent with the 
size of the errors due to uncertainties in the geometrical factors.

We have previously determined the actual hole concentration per Cu, $p$, 
for various La concentrations $x$, and have obtained an 
empirical relation. \cite{Hanaki}
In this paper we use this empirical relation to convert $x$ into $p$, 
which are listed in Table I.  Hereafter we mostly use the $p$ value 
in referring to the samples, because $p$ is much more intuitively 
understandable than $x$. 

\section{RESULTS}

\subsection{In-plane resistivity}

\begin{figure}
\includegraphics[width=8.5cm]{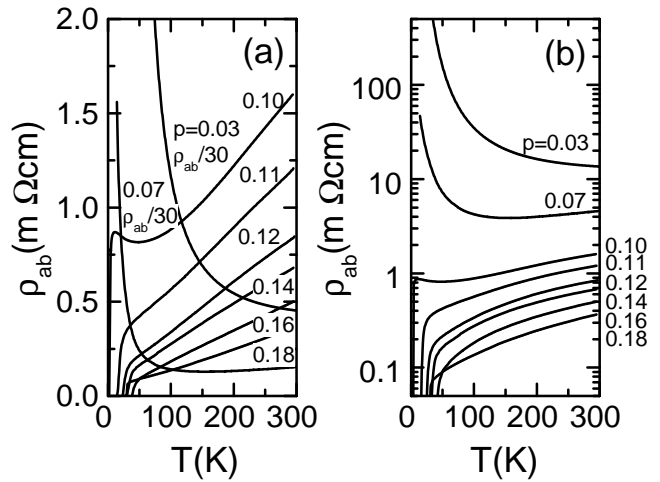}
\caption{(a) Temperature dependences of $\rho_{ab}$ of the BSLCO 
crystals for various $p$. (b) Semi-logarithmic plot of $\rho_{ab}$(T).}
\end{figure}

Figure 2 shows the temperature dependences of $\rho_{ab}$ for all the 
$p$ values studied here.  
As expected (and has been reported in Refs. \onlinecite{Murayama,Ono,Hanaki}), 
the magnitude of $\rho_{ab}(T)$ shows a systematic increase 
with decreasing carrier concentration from $p$ = 0.18 to 0.03. 
The $\rho_{ab}(T)$ curve of the overdoped sample ($p$ = 0.18) shows 
a behavior that can be described 
\cite{Murayama} by $a + bT^n$ with $n > 1$, 
while a strictly $T$-linear behavior is apparent in the 
optimally-doped sample ($p$ = 0.16).
In the most underdoped non-superconducting sample ($p$ = 0.03), 
$\rho_{ab}(T)$ shows 
an insulating behavior ($d\rho_{ab}/dT < 0$) from room temperature, 
which is quite different from the $\rho_{ab}(T)$ behavior of lightly 
hole-doped LSCO that shows a metallic behavior ($d\rho_{ab}/dT > 0$) 
even at $x$ = 0.01. \cite{mobility}  
We assert this difference is not intrinsic but is rather due to disorder, 
because the room-temperature sheet resistance per CuO$_2$ plane 
of our BSLCO at $p$ = 0.03 
is 110 k$\Omega$, while that of LSCO at $x$ = 0.03 
(Ref. \onlinecite{mobility}) is only 60 k$\Omega$; 
this indicates that a significant disorder is present 
in the heavily-underdoped BSLCO samples.  On the other hand, the residual 
resistivity at optimum-doping ($p$ = 0.16) is only 20 $\mu\Omega$cm, 
indicating that the electron transport can be quite clean in this system; 
since we find that the crystals are morphologically very clean 
for both $p$ = 0.03 and 0.16, the source of the electronic disorder is 
apparently not any crystalline defects but is likely to be the local 
disordering potentials due to the La dopants. 

It is useful to note that the barely superconducting sample, $p$ = 0.10, 
shows $\rho_{ab}$ value of $\sim$0.7 m$\Omega$cm just above $T_c$; 
this value corresponds to the sheet resistance per CuO$_2$ plane of 
5.8 k$\Omega$, which is very close to the quantum sheet resistance 
$h/4e^2$ (= 6.45 k$\Omega$).  This suggests that the disappearance of 
the superconductivity in BSLCO at unusually high doping of $p \simeq 0.10$ 
is caused by the disorder-driven superconductor-insulator transition in 
two-dimensional (2D) superconductors. \cite{Fukuzumi}

\subsection{Out-of-plane resistivity}

\begin{figure*}
\includegraphics[width=13cm]{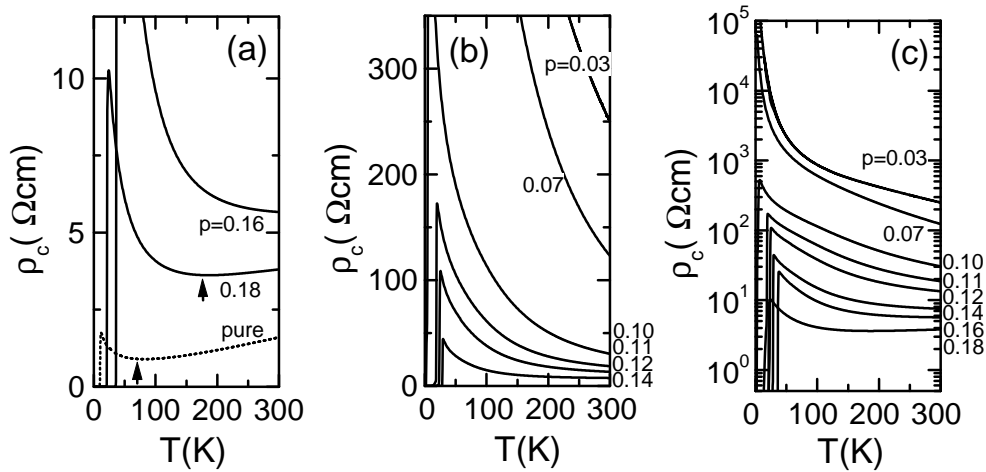}
\caption{Temperature dependences of $\rho_{c}$ 
of the BSLCO crystals for (a) $0.16 \le p \le 0.18$ and 
(b) $0.03 \le p \le $ 0.14. 
Panel (a) also includes $\rho_c(T)$ data of ``pure" sample 
(La-free Bi$_{2.13}$Sr$_{1.89}$CuO$_{6+\delta}$).  
Arrows mark the temperature where $\rho_{c}(T)$ 
shows a minimum. (c) Semi-logarithmic plot of $\rho_{c}$(T) for 
all the $p$ values.}
\end{figure*}

Figure 3 shows the temperature dependences of $\rho_{c}$ for the same 
compositions as those in Fig. 2.  
As is the case with $\rho_{ab}(T)$, the magnitude of $\rho_{c}(T)$ shows 
a systematic increase with decreasing carrier concentration.  
Since the magnitude of $\rho_{c}$ changes 
by 2 orders of magnitude from $p$ = 0.18 to 0.03, the data for 
the overdoped and optimally-doped samples are shown in Fig. 3(a), 
while those of the underdoped samples are shown in Fig. 3(b), 
both in linear scale; Fig. 3(c) plots $\rho_{c}(T)$ in semi-logarithmic 
scale to equally show the behavior of all the samples. 
In Fig. 3(a), $\rho_c(T)$ data for La-free 
Bi$_{2.13}$Sr$_{1.89}$CuO$_{6+\delta}$ (``pure" sample) 
are also shown for comparison (dotted line). 
It is worthwhile to note that, for each composition, the $\rho_{c}$ value 
of our crystals is generally smaller than that reported for BSLCO in the 
past. \cite{Martin,Wang} 

We note that a ``metallic" temperature dependence of $\rho_c$ is often 
observed, as is the case with the present experiment, in overdoped samples 
at high enough temperatures, \cite{Nakamura,Takenaka,Watanabe,Lavrov} 
which probably 
indicates that the in-plane scattering of the electrons is involved 
in determining the temperature dependence of $\rho_c$. \cite{Jayannavar}
In fact, recent $c$-axis magnetoresistance study by Hussey \textit{et al.} 
showed that $\rho_c$ of LSCO is strongly affected by the scattering 
within the planes. \cite{Hussey}  This suggests that looking at
$\rho_c/\rho_{ab}$ is useful for investigating
the intrinsic mechanisms of the unusual $c$-axis transport, because 
the in-plane scattering rate is roughly cancelled out in $\rho_c/\rho_{ab}$.

\subsection{Anisotropy ratio $\rho_c$/$\rho_{ab}$}

\begin{figure}
\includegraphics[width=7cm]{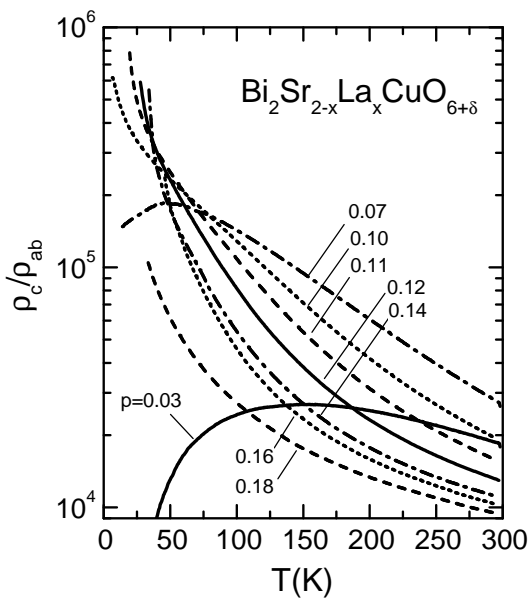}
\caption{Temperature dependences of $\rho_c$/$\rho_{ab}$ of the 
BSLCO crystals calculated from the data in Figs. 2 and 3.}
\end{figure}

Figure 4 shows the temperature dependences of the anisotropy ratio 
$\rho_c$/$\rho_{ab}$ for all the concentrations studied. 
At moderate temperatures, the magnitude of $\rho_{c}$/$\rho_{ab}$ 
monotonically increases with decreasing $p$, except for the 
most underdoped concentration ($p$ = 0.03) at which 
$\rho_{c}$/$\rho_{ab}$ suddenly drops; it is probably the case 
that this drop is caused by the heightened $\rho_{ab}$ value of the 
$p$ = 0.03 sample due to localization, and therefore if there were 
no enhanced disorder at this composition we would expect 
$\rho_{c}$/$\rho_{ab}$ to be not much different from that for $p$ = 0.07.
Note that the $\rho_c$/$\rho_{ab}$ value generally increases with 
lowering temperature, and almost reaches 10$^{6}$ just above $T_c$ 
for $p = 0.10 - 0.14$. 

Before our FZ-grown crystals became available, the anisotropic resistivity 
of BSLCO was measured by Wang \textit{et al.} using flux-grown crystals 
that showed rather large residual resistivities. \cite{Wang}
Though the data by Wang \textit{et al.} demonstrated reasonably 
systematic behavior and are mostly consistent with our data reported 
here, their data are different from ours in one important aspect: 
they observed that the anisotropy ratio $\rho_c/\rho_{ab}$ is almost 
independent of doping at 300 K, while we observed 
that $\rho_c/\rho_{ab}$ at 300 K increases systematically with 
decreasing $p$, as can be seen in Fig. 4. 

We note that in the anisotropic resistivity measurements 
using the flux-grown crystals, it was already recognized that the 
BSLCO system shows a very large anisotropy ratio 
$\rho_c/\rho_{ab}$ of more than 10$^5$, which is the largest among 
the cuprates; for example, Martin \textit{et al.} reported \cite{Martin} 
that $\rho_c/\rho_{ab}$ can be as large as 3$\times$10$^5$ in Bi-2201, 
and the work by Wang \textit{et al.} mentioned above reported \cite{Wang} 
maximum $\rho_c/\rho_{ab}$ of 2.5$\times$10$^5$. 
In our crystals, perhaps because of the smaller $\rho_{ab}$ values, 
the observed $\rho_c/\rho_{ab}$ is even larger: the $\rho_c/\rho_{ab}$ 
value exceeds 5$\times$10$^5$ in four of the concentrations 
($p$ = 0.10, 0.11, 0.12, and 0.14), and the maximum $\rho_c/\rho_{ab}$ 
we observe is 8$\times$10$^5$. 

\section{DISCUSSIONS}

\subsection{La-free (``pure") sample}

\begin{figure}
\includegraphics[width=6cm]{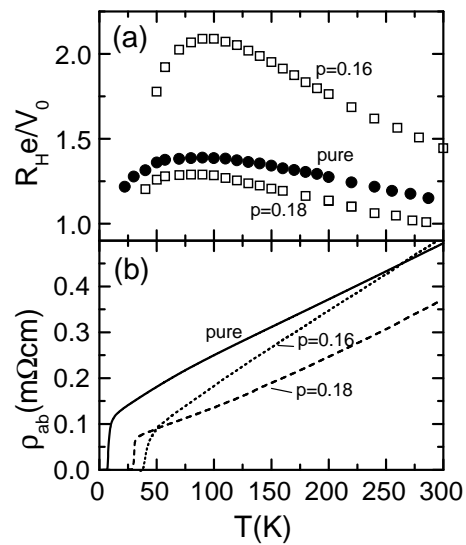}
\caption{(a) Plots of the renormalized Hall coefficient, $R_{H}e/V_0$, 
vs $T$ for the pure sample and the BSCLO samples with $p$ = 0.16 and 0.18, 
where $V_0$ is the volume per one Cu atom in the unit cell; 
the values of $R_{H}e/V_0$ of 
various cuprates near room temperature have been found to agree for the 
same $p$ value, giving a good tool to estimate $p$. \cite{Hanaki}  
(b) Comparison of the $\rho_{ab}(T)$ data of the pure sample 
to those of BSLCO with $p$ = 0.16 and 0.18.}
\end{figure}

In Fig. 3, one can infer that the onset of steeply insulating 
behavior in $\rho_c(T)$ moves to higher temperature with decreasing 
$p$ in the La-doped series of samples, which is consistent 
with the expected behavior of the pseudogap-opening temperature $T^*$. 
Although the data for pure sample in Fig. 3(a) appears to conform nicely 
to this trend, there is one issue that we need to elaborated on: 
When we estimate the actual hole concentration of the pure sample 
using the method we reported in Ref. \onlinecite{Hanaki}, the $p$ value 
of this pure sample is 0.17 [Fig. 5(a)]; the slope 
of $\rho_{ab}(T)$ also indicate that the $p$ value of the pure samples 
should be around 0.17 [Fig. 5(b)].  On the other hand, compared to the 
La-doped $p$ = 0.18 sample, the $\rho_c$ value of the pure sample 
is \textit{smaller} and the onset temperature of the insulating behavior 
is \textit{lower}, both of which would normally be expected for more 
\textit{overdoped} sample. 
Although this apparent puzzle looks problematic, it can be 
understood to be a result of a combination of extrinsic effects: 
First, since the constituent atoms of the apical-oxygen layer are 
different [(Sr,La)O for BSLCO and (Sr,Bi)O for pure 
samples], it is expected that the $c$-axis tunneling matrix element 
$t_{\perp}$ is also different; it appears that $t_{\perp}$ is larger 
in the pure samples, though the microscopic reason for the difference 
is not very clear. 
Second, since the $c$-axis magnetoresistance study of BSLCO 
has revealed \cite{Lavrov} that in the overdoped region the 
pseudogap is largely due to superconducting fluctuations and thus 
$T^*$ is governed primarily by $T_c$, the low $T_c$ value of the 
pure sample is the reason for the low onset temperature of the insulating 
behavior. \cite{note}

In Fig. 5(b), one notices that residual resistivity of the pure sample
is higher than that of the La-doped samples, which may be
counter-intuitive for a ``pure" sample. It is important to recognize,
however, that the composition of the ``pure" sample is
Bi$_{2.13}$Sr$_{1.89}$CuO$_{6+\delta}$, which means that there is sizable
disorder in the Sr-O block due to excess Bi. (It is known that
stoichiometric Bi$_{2.0}$Sr$_{2.0}$CuO$_{6+\delta}$ is unstable and cannot
be grown.) The larger residual resistivity and the lower $T_c$ of the
pure sample suggest that the disorder due to (Sr,Bi) substitution is
more harmful to the electronic system in this material than the
disorder due to (Sr,La) substitution.

\subsection{Two origins of the pseudogap}

It is worthwhile to mention that previous measurements of the $c$-axis
magnetoresistance of BSLCO have uncovered two additive mechanisms for
the pseudogap.\cite{Lavrov} In underdoped samples, a
magnetic-field-insensitive pseudogap emerges at $T^*$, and a distinct
one is formed above $T_c$ as a precursor to superconductivity (which is 
essentially the superconducting fluctuations); it was
found that the latter is mostly responsible for large negative
magnetoresistance in $\rho_c$. In overdoped samples, on the other hand,
it is only the precursor to superconductivity that appears above $T_c$
and serves as pseudogap.

Recently, scanning tunneling spectroscopy (STS) measurements have been
done on our pure samples (which is overdoped) and the pseudogap was
found to open below $68 \pm 2$ K. \cite{Kugler} In Fig. 3(a), the
minimum in $\rho_c(T)$ of the pure sample is observed at 70 K, which
actually agrees with $T^*$ determined by STS and gives strong support to
the interpretation that the steeply insulating behavior in $\rho_c(T)$
is largely due to the suppression of the DOS in BSLCO.
Moreover, the STS study has concluded \cite{Kugler} that the pseudogap
and the superconductivity has a common origin in our pure sample, which
is in agreement with the conclusion drawn from the magnetoresistance
study \cite{Lavrov} that the pseudogap in the overdoped region is in
fact a precursor to superconductivity.

\subsection{Temperature dependence of $\rho_c$}

\begin{figure}
\includegraphics[width=7cm]{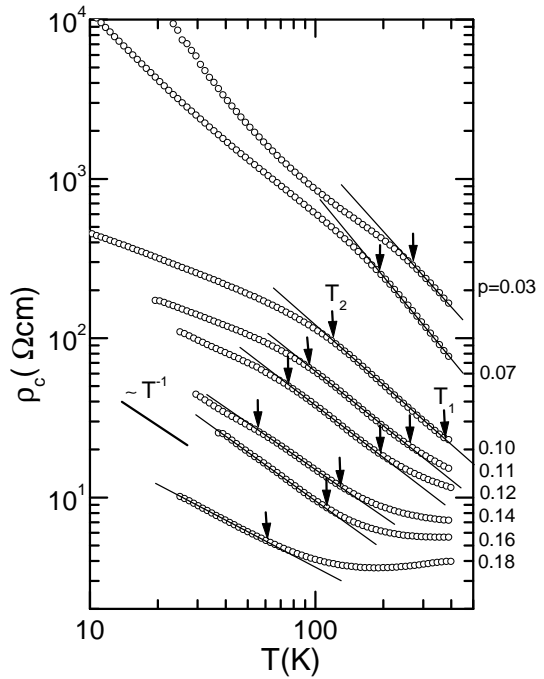}
\caption{Double-logarithmic plot of the temperature dependences of 
$\rho_c$ up to 400 K.
Thin solid lines are fits of the intermediate-temperature data to a 
$T^{-\alpha}$ dependence with $\alpha \simeq 1$. The slope for $\alpha$
= 1 is shown in the figure. Arrows mark $T_1$ and $T_2$, between which 
the $T^{-\alpha}$ dependence is observed.}
\end{figure}

During the course of the analysis of our $\rho_c(T)$ data, we found that
the log-log plot of the data (Fig. 6) uncovers intriguing systematics.
In Fig. 6, one can see that for most of the compositions there is a
finite range of temperature where the $\rho_c(T)$ data can be roughly
described by $T^{-\alpha}$ with $\alpha \simeq 1$ (thin solid lines in
Fig. 6). Since the pseudogap is known to cause a rapid increase in
$\rho_c$ with decreasing $T$, it is natural to interpret that the
observed $T^{-\alpha}$ dependence of $\rho_c$ is caused by the
development of the pseudogap; this interpretation motivates us to
evaluate the temperature $T_1$ below which the $\rho_c(T)$ obeys the
$T^{-\alpha}$ dependence. As shown in Fig. 7, the $p$-dependence of
$T_1$, as well as the value of $T_1$ itself, is consistent with what one
expects for the pseudogap temperature
$T^*$.\cite{Takenaka,Timusk,Watanabe,Lavrov}

\begin{figure}
\includegraphics[width=6cm]{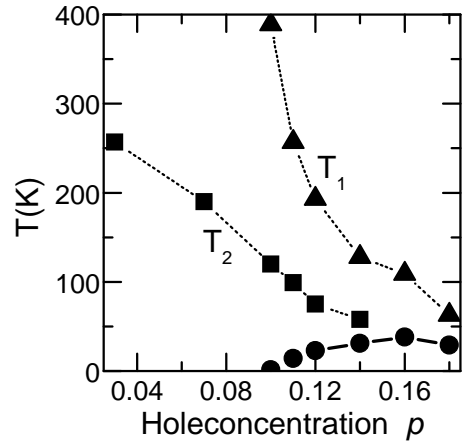}
\caption{$T$ vs $p$ diagram to show $T_1$ (our estimate of $T^*$ from 
the $\rho_c(T)$ data, triangles), $T_2$ (characteristic temperature for 
the saturation of the pseudogap development, squares), and $T_c$ (circles).}
\end{figure}

An interesting feature in the data in Fig. 6 is that the $\rho_c(T)$
data tend to deviate downwardly from the $T^{-\alpha}$ ($\alpha \simeq$ 1) 
dependence at
lower temperatures in underdoped samples; moreover, the characteristic
temperature for this deviation, $T_2$, systematically moves to higher
temperature with decreasing $p$. In the light of the current
understanding of the $c$-axis transport phenomena in cuprates, it is
most reasonable to interpret this deviation to come from a saturation of
the rapid pseudogap opening, and the more slower increase in $\rho_c$
below $T_2$ is likely to be due to the temperature dependence in the
strength of the confinement. Therefore, the temperature dependence of
$\rho_c$ as revealed in the log-log plot appears to reflect the fact
that the ``insulating" behavior is caused by the two phenomena, the
pseudogap and the confinement. This is new information which the
systematics of our data allows us to infer.
(The low-temperature data of the non-superconducting samples 
($p$ = 0.03 and 0.07), which show steeper divergence below $T_2$ 
compared to the superconducting samples, are probably affected by the
additional effect of localization.)

We note that the effect of superconducting fluctuations, which cause
additional pseudo-gapping near $T_c$,\cite{Lavrov} appears to be
inconsequential in the overall behavior of the underdoped samples shown
in Fig. 6, and the evolution of $T_2$ with $p$ is unrelated to $T_c$.
This is probably because the additional reduction in the DOS near $T_c$
due to superconducting fluctuations is relatively small in underdoped
samples, which can be inferred by the small size of the
magnetoresistance\cite{Lavrov} observed in underdoped samples.

It is useful to mention that several empirical temperature dependences 
have been proposed to describe the $\rho_c(T)$ data of cuprates.
In particular, $\rho_c(T) = (a/T)e^{(\Delta/T)} + bT + c$ has been 
reported to fit the $\rho_c(T)$ data of Bi-2212 very well, 
\cite{Watanabe97,Yan} while $\rho_c(T) = a_0 + aT + 1/(bT+c)$ has been  
reported to fit the $\rho_c(T)$ data of BSLCO. \cite{Wang}
We have tried to fit our data using these formulae, and found that 
the latter formula fits well to the data for $p$ = 0.16 and 0.18; 
however, neither of them fits well to the data for underdoped samples, 
which is apparently due to the existence of ``$T_2$" below which 
the divergence of $\rho_c(T)$ weakens.
Therefore, the extended range of doping of our study unveils the 
limitation of empirical temperature dependences.

\subsection{$\rho_c$/$\rho_{ab}$}

\begin{figure}
\includegraphics[width=7cm]{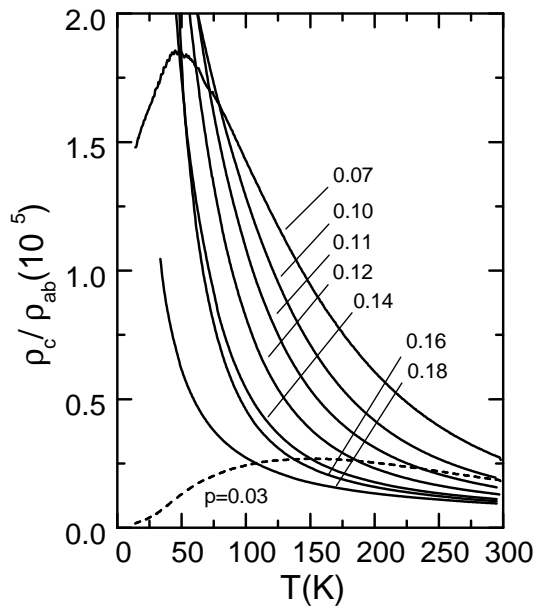}
\caption{Temperature dependences of $\rho_c$/$\rho_{ab}$ of the 
BSLCO crystals plotted in linear scale.}
\end{figure}

Figure 8 replots the $\rho_c/\rho_{ab}$ data of Fig. 4 
using a linear scale, which makes it easier to infer the quantitative 
change of the anisotropy with doping.  In Fig. 8, one can see that at 
high temperatures the $\rho_c/\rho_{ab}$ value drops rapidly with 
increasing $p$ for $p \ge 0.07$, but this rapid change tends to saturate 
in highly doped samples ($p \ge 0.14$); this situation can be understood 
more graphically by plotting the $p$ dependence of $\rho_c$/$\rho_{ab}$ at 
200 K and 300 K, as is shown in Fig. 9.  It can be seen in Fig. 9 
that at 200 K the $p$ dependence of $\rho_c/\rho_{ab}$ for high doping (
$p \ge 0.14$) is weak, while the $\rho_c/\rho_{ab}$ value 
steeply increases with decreasing $p$ for lower doping ($p \le 0.12$), 
except for $p$ = 0.03.  
It is most likely that the 
weak $p$ dependence at high doping reflects the change in the strength 
of the charge confinement, while the steep increase in the 
underdoped samples reflects the enhancement of the 
anisotropy due to the pseudogap.  In fact, the $T_1$ value we determined in 
Fig. 6 is near 200 K for $p$ = 0.12.
Therefore, the peculiar hole-doping 
dependence of the anisotropy is understood 
to be a combined result of the confinement and the pseudogap.

\begin{figure}
\includegraphics[width=6cm]{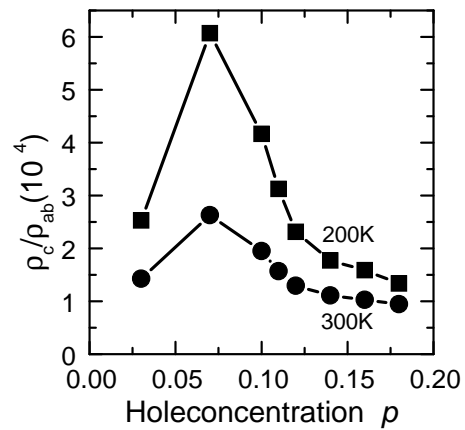}
\caption{$p$ dependences of $\rho_{c}$/$\rho_{ab}$ at 200 K and 300 K.}
\end{figure}

In Fig. 8, one notices that for non-superconducting concentrations  
($p$ = 0.03 and 0.07) there is a peak in the temperature 
dependence of $\rho_c/\rho_{ab}$ (at 50 and 100 K, respectively), 
and below the peak temperature the anisotropy is diminished. 
In particular, $\rho_c/\rho_{ab}$ for $p$ = 0.03 decreases dramatically 
with decreasing $T$, 
suggesting that the system is heading towards an anisotropic 
three-dimensional (3D) state for $T \rightarrow 0$. 
Such diminishment of $\rho_c/\rho_{ab}$ has also been observed in 
insulating samples of Bi-2212 (Ref. \onlinecite{Kitajima}) and 
LSCO (Ref. \onlinecite{Komiya}), 
and has been discussed to be due to localization of carriers.
In fact, when the carriers are completely localized and becomes 
immobile in \textit{any} direction, it is natural to see the system 
as a ``3D" insulator; note that the system is expected to retain 
some amount of anisotropy in the 3D-insulating state, reflecting the 
anisotropy of the ``bare" matrix elements.

\begin{figure}
\includegraphics[width=8cm]{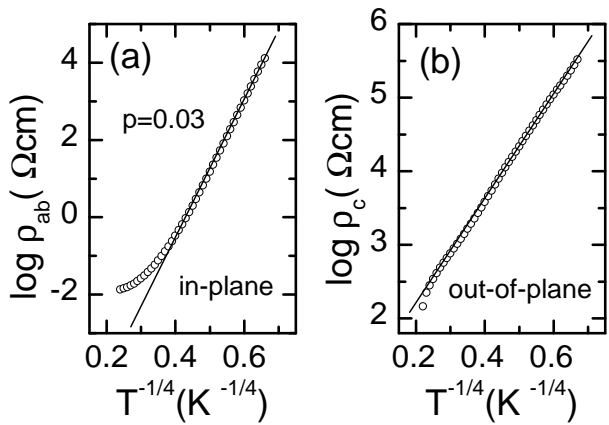}
\caption{(a,b) Log of $\rho_{ab}(T)$ and $\rho_{c}(T)$ of the BSLCO 
crystals at $p$ = 0.03 plotted vs $T^{-1/4}$. 
The dotted lines are fits of the data to the variable-range 
hopping dependence $\rho \sim \mathrm{exp}[(T_{0}/T)^{1/4}]$.}
\end{figure}

Using the resistivity data for $p$ = 0.03, where the superconductivity 
is lost and carriers are localized due to disorder, we can actually 
confirm that the system is approaching a 3D insulator. 
Figure 10 shows the 3D variable range hopping (VRH) 
plots of $\rho_{ab}(T)$ and $\rho_c(T)$ for $p$ = 0.03.  
The dotted lines are fits to the expected temperature dependence for 
the 3D VRH, $\rho \sim \mathrm{exp}[(T_{0}/T)^{1/4}]$. 
Since the 3D-VRH formula describes the temperature dependences of both 
$\rho_{ab}$ and $\rho_c$ very well, one can conclude that 
the charge transport along \textit{all} directions is governed by 
the \textit{same} mechanism, giving a rationale to call the system 
to be an anisotropic 3D insulator. 
The fits give $T_{0}$ values of $2.6 \times 10^6$ K and 
$8.6 \times 10^4$ K for $\rho_{ab}(T)$ and $\rho_c(T)$, respectively. 
These values are in the same range as the $T_0$ values obtained 
for flux-grown LSCO crystals in the lightly-doped regime, \cite{Chen} 
but the source of its anisotropy is not clear at this stage. 

\section{SUMMARY}

Systematic measurements of the $c$-axis resistivity $\rho_c$ and the
resistivity anisotropy ratio $\rho_c/\rho_{ab}$ of BSLCO for a wide
doping range showcase that the anisotropy is determined by three
mechanisms: charge confinement, pseudogap, and localization. At high
temperature and in highly doped samples, the charge confinement is the
sole determining factor, and thus one can measure the intrinsic strength
of the charge confinement in this corner of the phase diagram; our data
indicate that without the effect of the pseudogap the $\rho_c/\rho_{ab}$
value is order of 10$^4$ for BSLCO. At low temperature and in low doped
samples, the localization governs the charge transport and the
$\rho_c/\rho_{ab}$ ratio is diminished. In underdoped superconducting
samples, the pseudogap significantly enhances the $\rho_c/\rho_{ab}$
value with decreasing temperature, causing $\rho_c/\rho_{ab}$ to reach
$\sim$10$^6$. Close examination of the temperature dependence of
$\rho_c$ of underdoped samples finds that below $T^*$ there are two
distinct temperature regions where the divergence of $\rho_c$ appears to
be governed by different mechanisms; the data suggest that the
development of the pseudogap leads to a $T^{-\alpha}$ dependence of
$\rho_c$ with $\alpha \simeq$ 1 below $T^*$, while at lower temperature
$\rho_c$ shows a weaker temperature dependence where we argue that the
charge confinement dictates the $T$ dependence. Since the effect of the
pseudogap is apparently lacking in the anisotropic behavior of LSCO,
\cite{Komiya} the BSLCO system offers a complementary testing ground to
study the source of the peculiar $c$-axis transport in the cuprate
superconductors.

We thank A. N. Lavrov for helpful discussions.



\begin{thebibliography}{99}
\bibitem{note}
The BSLCO samples used in Ref. \onlinecite{Lavrov} were taken from the 
same batch as the samples used in the present study, but they were 
annealed with different conditions; this caused the $T_c$ values for 
the same composition to be lower in the samples of Ref. \onlinecite{Lavrov}.
This difference in $T_c$ (which affects $T^*$ in the overdoped 
region) is the reason for apparently different behaviors of $\rho_c(T)$ 
between the two studies.
\end{thebibliography}
\end{document}